# Utilizing Bochner's Theorem for Constrained Evaluation of Missing Fourier Data

**Jonathan Nemirovsky[1] and Efrat Shimron[2]**

[1] Physics Department and Solid State Institute, Technion – Israel Institute of Technology, Haifa 32000, Israel

[2] Biomedical Engineering Department, Technion – Israel Institute of Technology, Haifa 32000, Israel

E-mail: efrats@tx.technion.ac.il

**Abstract**

A method is presented for estimating unknown Fourier domain (k-space) data using a small number of samples in that space. The method is derived from Bochner's Theorem, and is termed: "Bochner Inequality Completion of K-Space (BICKS)". It is suitable for filling the k-space of a real and nonnegative unknown quantity, and applicable even when the sampling rate is substantially lower than the Nyquist sampling rate. The BICKS method is demonstrated in the context of medical imaging, but it is also applicable to many other scientific areas that utilize signal processing in Fourier domain. The results indicate that filling a highly undersampled k-space using BICKS enables high quality image reconstruction.

# 1. Introduction

The need to estimate unknown data in the Fourier domain arises in a variety of engineering branches. The acquired data is in many cases limited, therefore methods for completion of missing Fourier-domain data are concurrently sought.

In the context of medical imaging, several classical methods exploit properties of the Fourier domain for estimation of unknown data. Partial Fourier (PF) methods, employed in in Magnetic Resonance Imaging (MRI), utilize the Hermitian conjugate symmetry property of real-valued images for estimating data in half of k-space (Feinberg et al., 1986, Noll et al., 1991, Mcgibney et al., 1993). The method developed by Karp et al (1998) for Emission Computed Tomography utilizes Fourier domain constrains for filling gaps in the acquired data. The Projection Onto Convex Sets (POCS) approach, implemented in many fields including Computed Tomography (CT) and MRI, applies a priori known constraints in both the space and Fourier domains, and employs an iterative reconstruction process (Haacke et al., 1991, Sezan and Stark, 1984). The more general and powerful framework of *Compressed Sensing* (CS), which enables image reconstruction from limited samples, uses the sampled data as a constrain (Donoho, 2006, Candès et al., 2006); this approach has been proven useful for various modalities and applications (Lustig et al., 2007, Gamper et al., 2008, Candès et al., 2006, Trzasko and Manduca, 2009, Otazo et al., 2010)

In this paper, we present a method for estimating missing Fourier data using constrains which are derived from Bochner's Theorem (Katznelson, 1968, Bochner, 1933). This method, which we term Bochner Inequality Completion of K-Space (BICKS), is suitable for filling the k-space of a real and nonnegative quantity. Markedly, it is applicable even when the sampling rate is substantially low with respect to the Nyquist sampling rate. In the next section we present the theoretical derivation of the method, and discuss implementation aspects. Following the derivation, we demonstrate the method only in the context of medical imaging. Nevertheless, it is important to note that it may be applicable to other scientific areas in which the data is processed in the Fourier domain.

## 2. Theory

*2.1 Problem Definition*

Let $\rho(\mathbf{r}) \geq 0$ be the quantity that we seek to reconstruct, (e.g. an image). We assume that it is defined in a finite volume $0 \leq x, y, z \leq L$, where $\mathbf{r} = (x, y, z)$. For practical means it is useful to approximate $\rho(\mathbf{r})$ with a step-function defined on a grid of $N^3$ voxels of size $(L/N)^3$. The Discrete Fourier Transform (DFT) of this step-function density is defined by,

$$\hat{d}(\mathbf{k}_{ijk}) = \sum_{0 \leq i',j',k' < N} \exp(-i\mathbf{k}_{ijk} \cdot \mathbf{r}_{i'j'k'}) \rho(\mathbf{r}_{i'j'k'}), \qquad (1)$$

The Fourier coefficients $\hat{d}(\mathbf{k}_{ijk})$ are fundamentally acquired in medical imaging systems such as MRI. They are defined in this work on k-space (defined here as a discrete Cartesian grid, $\mathbf{k}_{ijk} \in \frac{2\pi}{L} Z_N^{\,3}$ where $0 \leq i, j, k < N$ and $Z_N := \{0, 1, 2, ..., N-1\}$).

The sought image $\rho(\mathbf{r}_{ijk})$ is reconstructed by the Inverse DFT (IDFT)

$$\rho(\mathbf{r}_{i'j'k'}) = \frac{1}{N^3} \sum_{0 \leq i,j,k < N} \exp(i\mathbf{k}_{ijk} \cdot \mathbf{r}_{i'j'k'}) \hat{d}(\mathbf{k}_{ijk}). \qquad (2)$$

In the following section we show that when the coefficients $\hat{d}(\mathbf{k}_{ijk})$ are undersampled, Bochner's theorem can be used to approximate some of the missing data in k-space. For this aim we assume that the reconstructed quantity is real and non-negative ($\rho(\mathbf{r}_{ijk}) = \rho_{ijk} \geq 0$).

*Bochner Inequality Completion of K-Space (BICKS) Theory*

By definition (Katznelson, 1968, Bochner, 1933), the coefficients $\hat{d}(\mathbf{k}_{ijk})$ are *positive definite* if for every choice of $n$ vectors $\xi_1, \xi_2, ..., \xi_n \in \frac{2\pi}{L} Z_N^{\,3}$ and every choice of $n$ complex numbers $z_1, z_2, ..., z_n \in C$ we have $\sum_{1 \leq i,j \leq n} z_i^* \hat{d}(\xi_i - \xi_j) z_j \geq 0$, with $\xi_i - \xi_j$ interpreted as a Modular Arithmetic notation in $\frac{2\pi}{L} Z_N^3$ (e.g., for $\xi_1 = \mathbf{k}_{i_1 j_1 k_1}$ and $\xi_2 = \mathbf{k}_{i_2 j_2 k_2}$ we define $\xi_1 - \xi_2 = \mathbf{k}_{i_1 j_1 k_1} - \mathbf{k}_{i_2 j_2 k_2} = \mathbf{k}_{i_3 j_3 k_3}$ with $i_3 = i_1 - i_2 \pmod{N}$, $j_3 = j_1 - j_2 \pmod{N}$ and $k_3 = k_1 - k_2 \pmod{N}$ ).

It is well known from Bochner's theorem (Katznelson, 1968, Bochner, 1933) that the coefficients $\hat{d}(k_{ijk})$ are positive definite if and only if $\rho(\mathbf{r}_{ijk}) = \rho_{ijk}$ are <u>non-negative</u> (i.e., $\rho(\mathbf{r}_{ijk}) = \rho_{ijk} \geq 0$ for every $0 \leq i, j, k < N$ ).

It is useful to define a normalized (dimensionless) non-negative density,

$$f(\mathbf{r}_{ijk}) := \frac{\rho(\mathbf{r}_{ijk})}{\sum_{ijk} \rho(\mathbf{r}_{ijk})} \geq 0 \tag{3}$$

The Discrete Fourier transform of $f(\mathbf{r}_{ijk})$ is given by,

$$\hat{f}(\mathbf{k}_{ijk}) := \frac{\hat{d}(\mathbf{k}_{ijk})}{\hat{d}(\mathbf{k}_{000})} \tag{4}$$

and it is easy to see that, $|\hat{f}(\mathbf{k}_{ijk})|^2 \leq |\hat{f}(\mathbf{k}_{000})|^2 = 1$ and $\hat{f}(\mathbf{k}_{ijk}) = \hat{f}(-\mathbf{k}_{ijk})^* = \hat{f}(\mathbf{k}_{-i-j-k})^*$, where "-" is interpreted here with the Modular Arithmetic notation $-m := N - m \pmod{N}$.

We note that $\hat{f}(\mathbf{k}_{ijk})$ is simply k-space normalized by the DC point value $\hat{d}(\mathbf{k}_{000})$.

From Bochner's Theorem for the case of $n = 3$, with the following choice of vectors

$$\xi_1 = \mathbf{k}_{i_1 j_1 k_1} = \frac{2\pi}{L}(i_1, j_1, k_1), \xi_2 = \mathbf{k}_{i_2 j_2 k_2} = \frac{2\pi}{L}(i_2, j_2, k_2) \text{ and } \xi_3 = \mathbf{k}_{000} = (0,0,0),$$

we conclude that the 3x3 matrix

$$\begin{bmatrix} 1 & \hat{f}(\mathbf{k}_{i_2 j_2 k_2} - \mathbf{k}_{i_1 j_1 k_1}) & \hat{f}(\mathbf{k}_{i_1 j_1 k_1}) \\ \hat{f}(\mathbf{k}_{i_2 j_2 k_2} - \mathbf{k}_{i_1 j_1 k_1})^* & 1 & \hat{f}(\mathbf{k}_{i_2 j_2 k_2}) \\ \hat{f}(\mathbf{k}_{i_1 j_1 k_1})^* & \hat{f}(\mathbf{k}_{i_2 j_2 k_2})^* & 1 \end{bmatrix} \tag{5}$$

is positive definite.

In particular it follows that,

$$\det \begin{bmatrix} 1 & \hat{f}(\mathbf{k}_{i_1 j_1 k_1}) & \hat{f}(\mathbf{k}_{i_2 j_2 k_2}) \\ \hat{f}(\mathbf{k}_{i_1 j_1 k_1})^* & 1 & \hat{f}(\mathbf{k}_{i_2 j_2 k_2} - \mathbf{k}_{i_1 j_1 k_1}) \\ \hat{f}(\mathbf{k}_{i_2 j_2 k_2})^* & \hat{f}(\mathbf{k}_{i_2 j_2 k_2} - \mathbf{k}_{i_1 j_1 k_1})^* & 1 \end{bmatrix} \geq 0 \quad (6)$$

This last inequality is equivalent to:

$$1 + 2\operatorname{Re}\left(\hat{f}(\mathbf{k}_{i_1 j_1 k_1}) \hat{f}(\mathbf{k}_{i_2 j_2 k_2})^* \hat{f}(\mathbf{k}_{i_2 j_2 k_2} - \mathbf{k}_{i_1 j_1 k_1})\right) - \\ \left(\left|\hat{f}(\mathbf{k}_{i_1 j_1 k_1})\right|^2 + \left|\hat{f}(\mathbf{k}_{i_2 j_2 k_2})\right|^2 + \left|\hat{f}(\mathbf{k}_{i_2 j_2 k_2} - \mathbf{k}_{i_1 j_1 k_1})\right|^2\right) \geq 0 \quad . \quad (7)$$

Abbreviating the notations with the following definitions $\hat{f}_1 := \hat{f}(\mathbf{k}_{i_1 j_1 k_1})$, $\hat{f}_2 := \hat{f}(\mathbf{k}_{i_2 j_2 k_2})$, $\hat{f}_{12} := \hat{f}(\mathbf{k}_{i_2 j_2 k_2} - \mathbf{k}_{i_1 j_1 k_1})$ and rearranging terms, we obtain

$$1 - \left(\left|\hat{f}_1\right|^2 + \left|\hat{f}_2\right|^2\right) \geq \left[\operatorname{Re}(\hat{f}_{12})\right]^2 + \left[\operatorname{Im}(\hat{f}_{12})\right]^2 - 2\operatorname{Re}(\hat{f}_1^* \hat{f}_2)\operatorname{Re}(\hat{f}_{12}) - 2\operatorname{Im}(\hat{f}_1^* \hat{f}_2)\operatorname{Im}(\hat{f}_{12}) \quad . \quad (8)$$

Rearranging terms again we obtain

$$1 - \left(\left|\hat{f}_1\right|^2 + \left|\hat{f}_2\right|^2\right) \geq \left[\operatorname{Re}(\hat{f}_{12}) - \operatorname{Re}(\hat{f}_1^* \hat{f}_2)\right]^2 + \left[\operatorname{Im}(\hat{f}_{12}) - \operatorname{Im}(\hat{f}_1^* \hat{f}_2)\right]^2 - \left[\operatorname{Re}(\hat{f}_1^* \hat{f}_2)\right]^2 - \left[\operatorname{Im}(\hat{f}_1^* \hat{f}_2)\right]^2$$

$$(9)$$

or simply

$$1 - \left(\left|\hat{f}_1\right|^2 + \left|\hat{f}_2\right|^2\right) \geq \left|\hat{f}_{12} - \hat{f}_1^* \hat{f}_2\right|^2 - \left|\hat{f}_1^* \hat{f}_2\right|^2 \quad . \quad (10)$$

We conclude that

$$1 - \left(\left|\hat{f}(\mathbf{k}_{i_1 j_1 k_1})\right|^2 + \left|\hat{f}(\mathbf{k}_{i_2 j_2 k_2})\right|^2\right) + \left|\hat{f}(\mathbf{k}_{i_1 j_1 k_1})^* \hat{f}(\mathbf{k}_{i_2 j_2 k_2})\right|^2 \geq \left|\hat{f}(\mathbf{k}_{i_2 j_2 k_2} - \mathbf{k}_{i_1 j_1 k_1}) - \hat{f}(\mathbf{k}_{i_1 j_1 k_1})^* \hat{f}(\mathbf{k}_{i_2 j_2 k_2})\right|^2$$

$$(11)$$

or simply

$$\left(1 - \left|\hat{f}(\mathbf{k}_{i_1 j_1 k_1})\right|^2\right)\left(1 - \left|\hat{f}(\mathbf{k}_{i_2 j_2 k_2})\right|^2\right) \geq \left|\hat{f}(\mathbf{k}_{i_2 j_2 k_2} - \mathbf{k}_{i_1 j_1 k_1}) - \hat{f}(\mathbf{k}_{i_1 j_1 k_1})^* \hat{f}(\mathbf{k}_{i_2 j_2 k_2})\right|^2 \quad . \quad (12)$$

Substituting $\mathbf{p}_1 = -\mathbf{k}_{i_1 j_1 k_1}$, $\mathbf{p}_2 = \mathbf{k}_{i_2 j_2 k_2}$ and using $\hat{f}(-\mathbf{p}_1)^* = \hat{f}(\mathbf{p}_1)$ we finally conclude that

$$\left(1 - \left|\hat{f}(\mathbf{p}_1)\right|^2\right)\left(1 - \left|\hat{f}(\mathbf{p}_2)\right|^2\right) \geq \left|\hat{f}(\mathbf{p}_1 + \mathbf{p}_2) - \hat{f}(\mathbf{p}_1)\hat{f}(\mathbf{p}_2)\right|^2 \quad (13)$$

for every $\mathbf{p}_1, \mathbf{p}_2 \in \frac{2\pi}{L} Z_N^3$.

We note that $\mathbf{p}_1$ and $\mathbf{p}_2$ are vectors that stretch from the center of k-space (i.e. the DC point) to the two pixels positions (see figure 1a).

Since $\hat{f}(\mathbf{p}_1)$, $\hat{f}(\mathbf{p}_2)$, and $\hat{f}(\mathbf{p}_1 + \mathbf{p}_2)$ have complex values, let us interpret equation (13) in the complex plane, in which the axes correspond to real and imaginary values. Importantly, it follows from (13) that if $\hat{f}(\mathbf{p}_1)$ and $\hat{f}(\mathbf{p}_2)$ are known, the value of $\hat{f}(\mathbf{p}_1 + \mathbf{p}_2)$ must be inside a specific circle in the complex plane (figure 1b); the circle center is at $\hat{f}(\mathbf{p}_1)\hat{f}(\mathbf{p}_2)$, and its radius is

$$R(\mathbf{p}_1, \mathbf{p}_2) = \sqrt{\left(1 - |\hat{f}(\mathbf{p}_1)|^2\right)\left(1 - |\hat{f}(\mathbf{p}_2)|^2\right)} \quad . \tag{14}$$

We denote this circle by,

$$\text{Circ}_{\mathbf{p}_1, \mathbf{p}_2} := \left\{ (a + b\mathrm{i}) \;\middle|\; \left|(a + b\mathrm{i}) - \hat{f}(\mathbf{p}_1)\hat{f}(\mathbf{p}_2)\right| \leq R(\mathbf{p}_1, \mathbf{p}_2) \right\} \quad . \tag{15}$$

Markedly, this means that if two k-space pixels at positions $\mathbf{p}_1$ and $\mathbf{p}_2$ are sampled during an MRI Hydrogen density scan, the value of a third *unsampled* pixel located at $\mathbf{p}_{unsampled} = \mathbf{p}_1 + \mathbf{p}_2$ must be confined in the circle defined above; any point inside this circle is a legitimate estimate for $\hat{f}(\mathbf{p}_{unsampled})$. In particular, we estimate this unsampled value as the circle center, i.e. $\hat{f}(\mathbf{p}_1)\hat{f}(\mathbf{p}_2)$; for this choice the maximal error is $R(\mathbf{p}_1, \mathbf{p}_2)$.

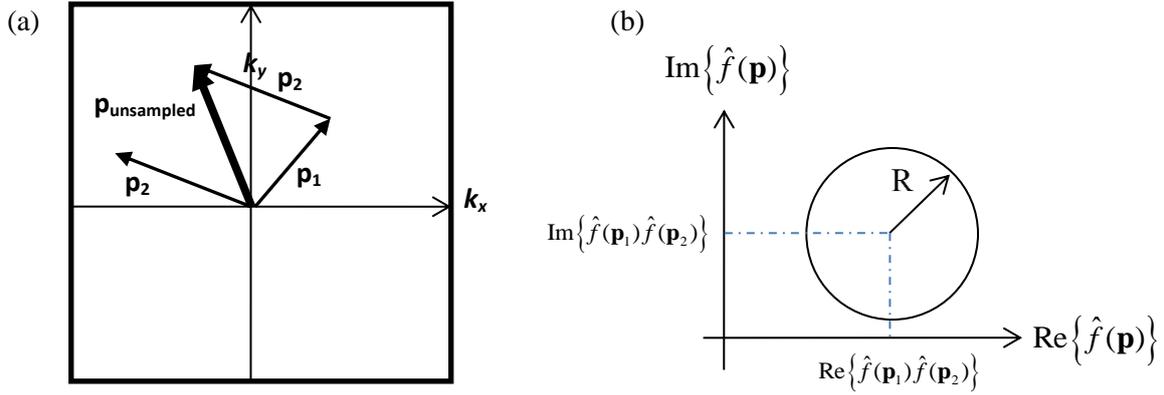

**Figure 1.** Geometrical interpretation of equation (13). (a) The k-space vectors $\mathbf{p}_1$ and $\mathbf{p}_2$ point to sampled pixels, and their sum points to an unsampled pixel $\mathbf{p}_{unsampled} = \mathbf{p}_1 + \mathbf{p}_2$. (b) In the complex plane, the circle created according to (13) is used for evaluation of $\hat{f}(\mathbf{p}_{unsampled})$.

When there are multiple pairs of known pixels that contribute to the evaluation of a single unknown pixel, the unknown value must lie within the intersection of all the associated circles. Consider a k-space pixel that can be reached by $n$ different sampled pixel pairs, $\mathbf{p}_{unsampled} = \mathbf{p}_j + \mathbf{q}_j$ $(j = 1, 2, 3, ..., n)$. The intersection area of the $n$ circles associated with these pairs is defined as $S_{\mathbf{p}_{unsampled}} := \bigcap_{j=1}^{n} \text{Circ}_{\mathbf{p}_j, \mathbf{q}_j}$, and $\hat{f}(\mathbf{p}_{unsampled})$ must lie inside this convex area, i.e., $\hat{f}(\mathbf{p}_{unsampled}) \in S_{\mathbf{p}_{unsampled}}$. Clearly, any point in $S_{\mathbf{p}_{unsampled}}$ is a legitimate estimate for $\hat{f}(\mathbf{p}_{unsampled})$. Intuitively, the simplest estimate for $\hat{f}(\mathbf{p}_{unsampled})$ is the center of mass value of this convex intersection area.

As an example, consider the case $\mathbf{p}_{unsampled} = \mathbf{p}_1 + \mathbf{p}_2 = \mathbf{p}_3 + \mathbf{p}_4 = \mathbf{p}_5 + \mathbf{p}_6$, where pixels 1 to 6 are known (figure 2). The three pairs of vectors can be used to estimate the pixel at position $\mathbf{p}_{unsampled}$. From (13) it is known that $\hat{f}(\mathbf{p}_{unsampled})$ must be inside the three circles associated with the three pairs. Hence, a reasonable estimate for this unknown value is the intersection area center of mass. Let

Utilizing Bochner's Theorem for Constrained Evaluation of Missing Fourier Data

us denote the coordinates of this center of mass by $(a,b)$; the unknown pixel therefore receives the value $\hat{f}(\mathbf{p}_{unsampled}) = a + b\mathrm{i}$, where $\mathrm{i} = \sqrt{-1}$ is the imaginary unit.

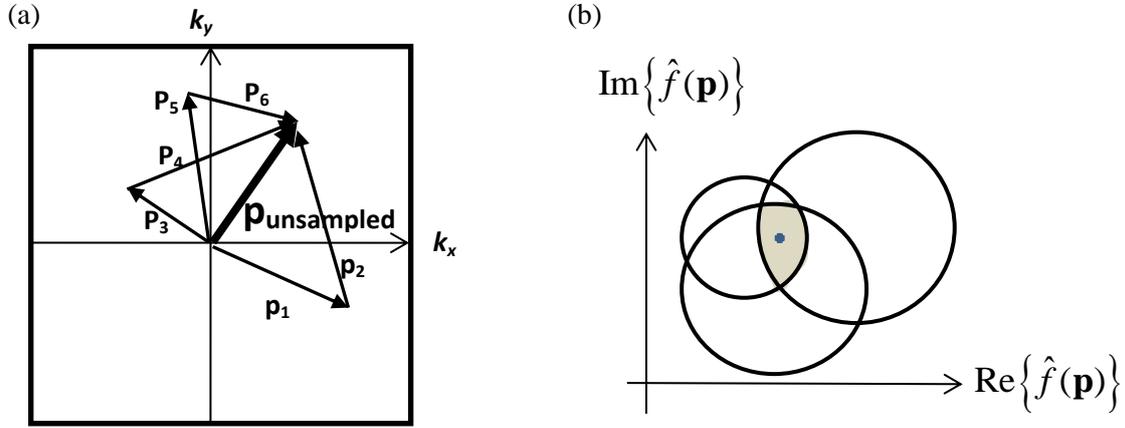

**Figure 2.** Evaluation of $\hat{f}(\mathbf{p}_{unsampled})$ using three circles. (a) In k-space, three pairs of sampled pixels point to $\mathbf{p}_{unsampled}$. (b) In the complex plane, $\hat{f}(\mathbf{p}_{unsampled})$ is evaluated using the three corresponding circles. It follows from (13) that $\hat{f}(\mathbf{p}_{unsampled})$ must be inside the intersection area (gray area); the cross marks the center of mass of that area.

*2.2 Error Estimation*

We now turn to evaluate the estimation error for a single pixel and for the whole image. Let $d\left(S_{\mathbf{p}_{unsampled}}\right)$ be the maximal distance between the circles intersection area center of mass and its boundary $\partial S_{\mathbf{p}_{unsampled}}$. It follows from eqs. (13-15) that the maximal error of estimating a single pixel in k-space is $err(\hat{f}(\mathbf{P}_{unsampled})) \leq d(S_{\mathbf{P}_{unsampled}})$. Generally, the circles centers do not coincide, hence

$$err(\hat{f}(\mathbf{p}_{unsampled})) << \min_{j} R(\mathbf{p}_j, \mathbf{q}_j) \ . \tag{16}$$

From (14) we see that $R(\mathbf{p}_j, \mathbf{q}_j)$ is small if $|f(\mathbf{p}_j)|^2$ or $|f(\mathbf{q}_j)|^2$ is close to 1. Therefore, the error is small if for one $j$ or more in the right hand side of (16) we have $|f(\mathbf{p}_j)|^2$ or $|f(\mathbf{q}_j)|^2$ close to 1.

Since pixels with $\hat{f}(\mathbf{p}) \to 1$ are found at the neighborhood of $\mathbf{p} = 0$, i.e. k-space origin, it follows that using such pixels for evaluation of unknown pixels reduces the evaluation error.

To estimate the error for the whole image, we utilize (16) and Plancharel's Theorem; it follows that the total image reconstruction error in the image domain is

$$|\text{error}|^2 := \sum_{ijk} \left| f_{ijk}^{\text{True}} - f_{ijk}^{\text{Estimated}} \right|^2 \leq \sum_{\mathbf{p}_{\text{unsampled}}} \left| d(S_{\mathbf{p}_{\text{unsampled}}}) \right|^2. \tag{17}$$

*2.3 Sampling Strategy*

Generally, the BICKS method is suitable for filling a k-space sampled with *any* scheme. However, to fully benefit from this method, two considerations must be addressed:

1. The sampled pixels should enable BICKS evaluation of all of the unsampled k-space pixels.
2. As discussed in the previous section, the error is reduced by sampling pixels with $\left|\hat{f}(\mathbf{p})\right|$ close to 1, which are usually located near the k-space origin.

A useful sampling strategy for the proposed BICKS method therefore includes dense sampling around k-space center, and sparse sampling at the rest of k-space. For example, in this work we use a fine rectilinear grid of sampled pixels near the origin, while the rest of k-space is sampled with a coarse grid (figure 3). Note that for this sampling scheme both 1 and 2 criterions mentions above are achieved.

Additionally, we found it useful to interpolate k-space data by zero-padding the image in the space domain; this increases the number of pixels with $\left|\hat{f}(\mathbf{p})\right|$ close to 1, thus it improves the estimation of unsampled pixels.

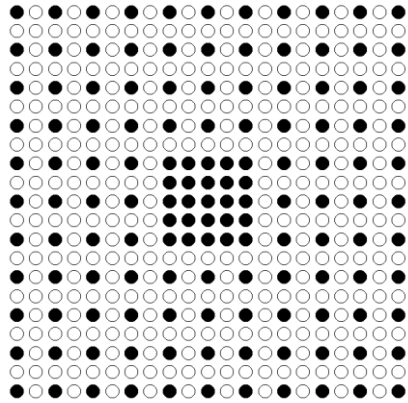

**Figure 3.** A useful k-space sampling scheme for BICKS evaluation of unsampled data. Black (white) dots represent sampled (unsampled) pixels.

*2.4 Image Reconstruction*

After k-space data is filled, the image can be reconstructed using the IDFT as in (2).

### 3. Methods

To demonstrate the applicability of the proposed BICKS method we implemented it on the general Shepp-Logan brain phantom (Shepp and Logan, 1974) and on a single-slice brain image obtained from an MRI scan (obtained with permission from http://www.eecs.berkeley.edu/~mlustig/Software.html). Although we note that the BICKS method is generally suitable for filling a 3D k-space, as described in the theory chapter; for simplicity, the demonstration given here includes a 2D implementation. All the numerical calculations were performed with MATLAB (Mathworks, Natick, MA, USA) on a computer with Intel(R) Core™2 Duo processor and 2.93GHz memory.

*3.1 Preparations*

Each image was first zero-padded externally to double its dimensions (for reasons discussed above), and then Fourier-transformed to obtain its full k-space. Then, the k-space was normalized so that its DC point equaled unity. Later, an image was reconstructed from this fully sampled normalized k-space using (2); this image is referred to as the 'gold standard'.

*3.2 Partial Sampling of* k-space

Following the BICKS error estimation and sampling strategy analysis given above, we applied a k-space sampling scheme that consisted of two parts: (i) dense sampling of k-space center, which included all pixels within a square of $N_D \times N_D$ pixels around the DC pixel, and (ii) sparse sampling, which included pixels on every other row and every other column, of the rest of k-space (figure 3).

In practice, we have implemented this scheme for partially sampling only *one half* of *k-space*; we then completed the corresponding partial data on the other half using the Conjugate Symmetry property of k-space. This method reduces the acquired data amount; hence, it increases the acceleration factor, defined as $R = N_{k-space} / N_{sampled}$, where $N_{k-space}$ is the total number of k-space pixels and $N_{sampled}$ is the number of sampled pixels. This sampling scheme was applied with $N_D = 0.035 \cdot N$, so that the central dense-sampling area was very small. In total, 12.5% of k-space pixels were sampled, therefore the acceleration factor was $R \approx 8$.

*3.3 BICKS Implementation*

The missing k-space data were evaluated using BICKS, where for each unsampled pixel we found several pairs of supporting sampled pixels so that $\mathbf{p}_{unsampled} = \mathbf{p}_j + \mathbf{q}_j$ $(j = 1, 2, 3, ..., n)$. Particularly, we selected only pairs in which one pixel was close to the DC point, because (as discussed above) using high-valued pixels improves the estimation. The other pixel was within a neighborhood of 3x3 pixels around $\mathbf{p}_{unsampled}$. Next, we calculated the corresponding circles, computed their intersection area center and assigned the corresponding value to the unsampled pixel.

Using this method, the entire unsampled k-space was filled with the estimated values. For comparison, we also generated the *zero-filled* k-space by endowing all the unsampled pixels with the zero value. The comparison did not include other Partial Fourier methods such as Homodyne and POCS (Noll et al., 1991, Haacke et al., 1991) since these methods cannot offer an acceleration factor higher than 2 (Liu et al., 2012), while in this work the acceleration factor was 8.

Once all the unsampled k-space pixels were filled, we reconstructed an image using the IDFT. We then removed the external zero padding for comparison with the original image.

*3.4 Image Analysis*

To quantify the difference between the original and reconstructed images, we calculated the Root Mean Square (RMS) measure, $RMS = \sqrt{\frac{\sum_{1 \leq i,j < N} |I_{ij}^{\text{Rec.}} - I_{ij}^{0}|^2}{N^2}}$, where $I_{ij}^{0}$ is the gold standard image reconstructed from the fully sampled k-space, $I_{ij}^{\text{Rec.}}$ is the image reconstructed from the partially sampled k-space, and $N^2$ is the total number of image pixels. To further assess the images similarity we calculated Pearson's Correlation Coefficient (CC),

$CC = \frac{\sum_{i,j}(I_{ij}^{\text{Rec.}} - I_{\text{Mean}}^{\text{Rec.}})(I_{ij}^{0} - I_{\text{Mean}}^{0})}{\sqrt{\sum_{i,j}(I_{ij}^{\text{Rec.}} - I_{\text{Mean}}^{\text{Rec.}})^2} \sqrt{\sum_{i,j}(I_{ij}^{0} - I_{\text{Mean}}^{0})^2}}$ where $I_{ij}^{\text{Rec.}}$, $I_{ij}^{0}$ are the intensities at the $ij^{th}$ pixel, $I_{\text{Mean}}^{\text{Rec.}}$ is the mean intensity of the reconstructed image and $I_{\text{Mean}}^{0}$ is the mean intensity of the gold standard image.

**4. Results**

We simulated partial k-space sampling with an 8-fold acceleration, filled the unsampled pixels with the proposed BICKS method, and calculated the reconstructed image. For comparison, we also filled the undersampled k-space with zeros and reconstructed an image.

For the Shepp-Logan example (figure 4), evaluating the unsampled pixels using BICKS rather than zeros significantly enhanced the k-space similarity to the fully sampled one. Consequently, the BICKS-reconstructed image had better quality than the zero-filling one; it contained sharper edges and clearer fine details. This was quantified by an improvement of 35% in the RMS measure and 12% in the CC measure (Table 1).

For the brain MRI example (figure 5), applying the BICKS estimation instead of zeros to the unsampled k-space pixels improved substantially the fidelity of the data in this domain, especially for the high frequencies. As a result, the BICKS-reconstructed image had better resolution and higher diagnostic value than the zero-filled reconstruction. This was quantified by a 38% improvement of the RMS measure and 4% improvement of the CC measure (Table 1). To summarize, the results indicate that estimating unsampled k-space data with the BICKS method is highly beneficial.

|  | Shepp-Logan image | | Brain MR image | |
|---|---|---|---|---|
| Zero-fill | RMS = 0.132 | CC=0.836 | RMS = 0.029 | CC=0.943 |
| BICKS | RMS = 0.087 | CC=0.936 | RMS = 0.018 | CC=0.977 |

**Table 1.** RMS and CC measures corresponding to the reconstructed images in figures 4 and 5.

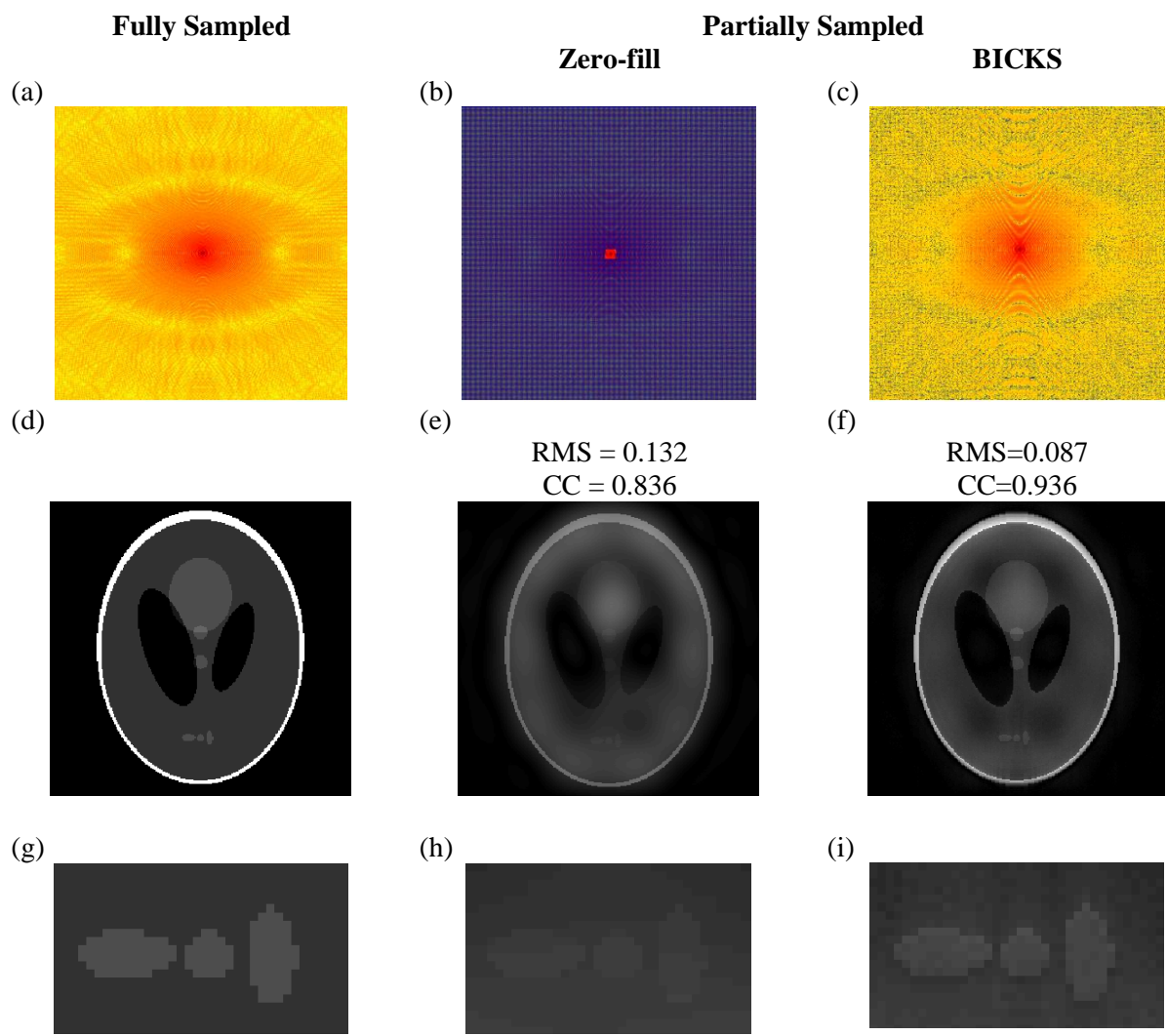

**Figure 4.** Shepp-Logan phantom (Shepp and Logan, 1974) reconstruction from an 8-fold undersampled k-space. **a** Fully sampled k-space. **b** Sampled k-space, with zero filling. **c** BICKS-filled k-space. In **a,b,c** the values are displayed in a logarithmic scale. **d:** The gold standard image. **e,f:** Reconstructions from the zero-filled and BICKS-filled k-spaces, respectively. **g,h,i:** Enlargement of the high-frequency details area in **d,e,f,** respectively.

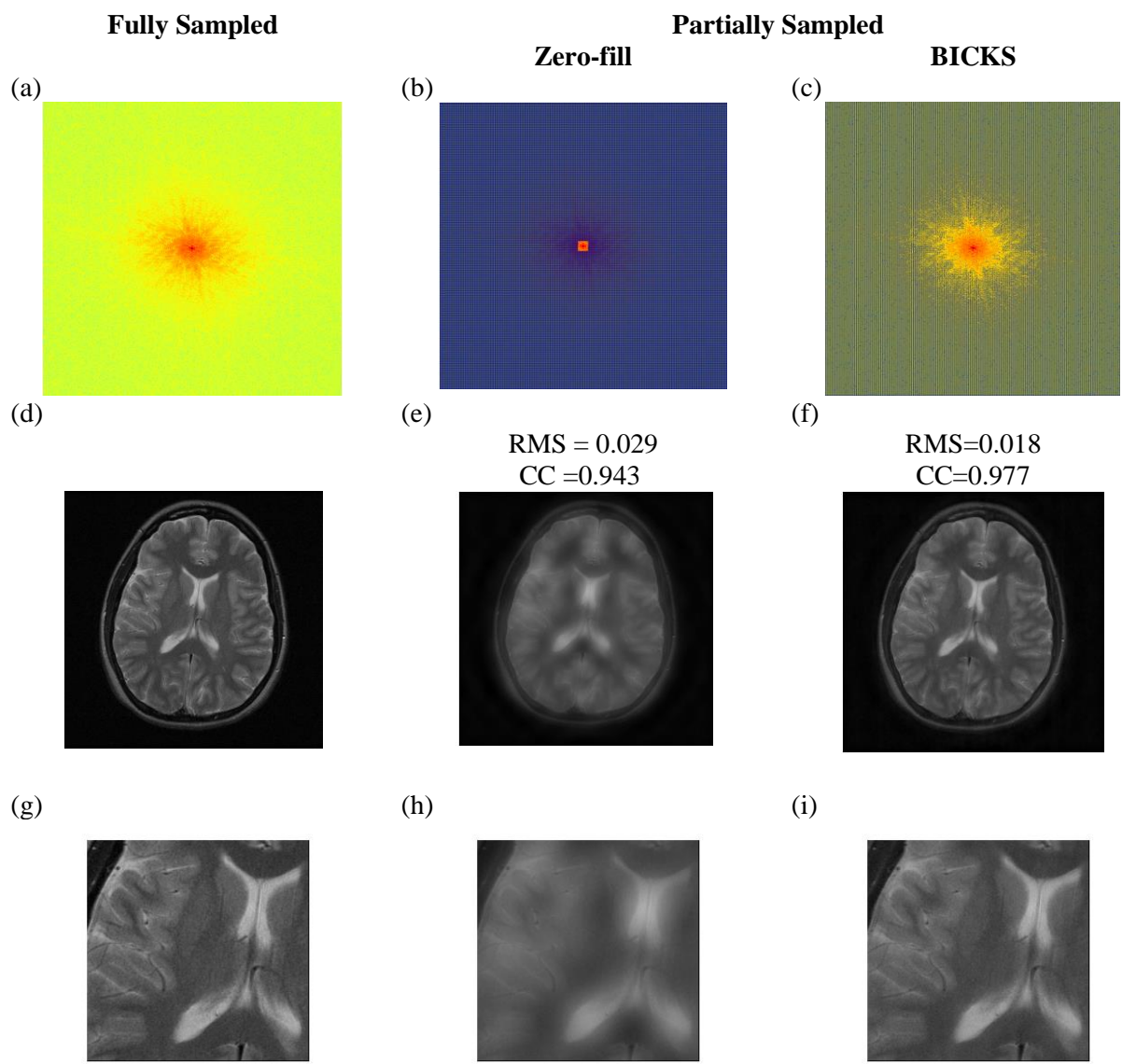

**Figure 5.** MR Brain image reconstruction from an 8-fold undersampled k-space. The brain image was adopted with permission from (Lustig et al., 2007) and obtained from http://www.eecs.berkeley.edu/~mlustig/Software.html .**a** Fully sampled k-space (logarithmic scale) **b,c** Sampled k-space, with zero filling and BICKS-filling respectively (logarithmic scale). **d** Gold standard image. **e,f** Reconstructions from the zero-filled and BICKS-filled k-spaces, respectively. **g,h,i** Enlargement of the central area in **d,e,f,** respectively.

## 5. Discussion

A novel method, entitled BICKS, is introduced. This method utilizes implications of Bochner's Theorem and k-space Hermitian Symmetry for evaluation of unsampled k-space data. We described the method derivation analytically, provided error estimation, and discussed the principles of an adequate k-space sampling scheme design.

The feasibility of implementing this method for medical imaging was demonstrated through simulations of a Shepp-Logan phantom and a single-slice brain MR image. The results demonstrated that filling a partially sampled k-space with BICKS instead of zeros results in higher quality images; the resolution enhances due to the faithful BICKS estimation of unsampled high-frequency k-space data.

Advantageously, the BICKS method does not restrict the partial-sampling ratio to any theoretical minimum. In our implementation, this method was combined with Conjugate Synthesis to enable high acceleration of MRI data acquisition. As demonstrated, qualitative images were reconstructed from k-spaces undersampled with an 8-fold acceleration. This acceleration is significantly higher than that enabled by PF methods, which is slightly less than 2 (Liu et al., 2012), and comparable to CS acceleration (Lustig et al., 2007). However, Compressed Sensing algorithms require extensive computations (Murphy et al., 2012, Ramani and Fessler, 2010b), while the proposed method involves non-iterative robust calculations.

Another appeal of the BICKS method is its applicability to various k-space sampling schemes. The current work focused on an ordered sampling strategy, which is highly effective for BICKS completion, and compatible with rectilinear MR imaging protocols. However, the method is applicable to other sampling schemes, which may contain one-dimensional or even multi-dimensional random features. The sampling scheme optimization is beyond the scope of this paper, and remains for future work.

In conclusion, the proposed BICKS method offers an effective completion of unacquired k-space data, sampling scheme flexibility and simple computations.